\documentclass{article}

\usepackage{arxiv}

\usepackage[utf8]{inputenc}
\usepackage[T1]{fontenc}
\usepackage{url}
\usepackage{booktabs}
\usepackage{amsfonts}
\usepackage{nicefrac}
\usepackage{microtype}
\usepackage{lipsum}
\usepackage{graphicx}
\usepackage{natbib}
\usepackage{doi}
\usepackage{orcidlink}
\usepackage{hyperref}

\newcommand{\orcidicon}[1]{%
  \href{https://orcid.org/#1}{%
    \raisebox{-0.15ex}{\includegraphics[height=2ex]{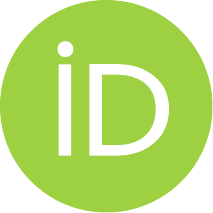}}%
  }%
}

\hypersetup{
  hidelinks,
  pdftitle={Machine Learning-based detection of long COVID using Heart Rate Variability Analysis},
  pdfsubject={},
  pdfauthor={Brais Iglesias-Otero},
  pdfkeywords={Long COVID, Heart Rate Variability, Machine Learning}
}

\title{Machine Learning-based detection of long COVID using Heart Rate Variability Analysis}

\author{
\normalfont
B. Iglesias-Otero$^{1,*}$ \orcidicon{0009-0000-8195-8234}
\quad
Xosé A. Vilar Sobrino$^{2}$ \orcidicon{0000-0002-8572-1637}
\quad
María J. Lado$^{2}$ \orcidicon{0000-0002-9027-7788}
\\[0.35em]
Leandro Rodríguez-Liñares$^{2}$ \orcidicon{0000-0002-7087-899X}
\quad
Baltasar García Pérez-Schofield$^{2}$ \orcidicon{0000-0002-0603-2842}
\quad
Pedro Cuesta-Morales$^{2}$ \orcidicon{0000-0003-2523-923X}
\\[0.35em]
Arturo J. Méndez$^{2}$ \orcidicon{0000-0002-0149-5581}
\quad
María Bustillo Casado$^{3}$ \orcidicon{0000-0002-0754-7637}
\quad
Alexandre García-Caballero$^{4}$ \orcidicon{0000-0001-5714-8266}
\\[0.9em]
\\
$^{2}$Information Processing and Telecommunications Center, Universidad Politécnica de Madrid,
ETSI Telecomunicación.
\\
$^{2}$Escola Superior de Enxeñaría Informática, IFCAE, Universidade de Vigo.
\\
$^{3}$Servicio de Medicina Interna, Complexo Hospitalario Universitario de Ourense
\\
$^{4}$Instituto Biomédico Galicia Sur, Centro de Investigación Biomédica en Red de Salud Mental (CIBERSAM)
\\[0.6em]
$^{*}$\texttt{brais.iglesias@upm.es}
}

\date{}

\renewcommand{\headeright}{}
\renewcommand{\undertitle}{}
\renewcommand{\shorttitle}{Machine Learning-based detection of long COVID using Heart Rate Variability Analysis}

\begin{document}

\maketitle

\begin{abstract}
After COVID epidemic has ravaged the world, around 20\% of infected subjects continue to manifest symptoms several months after their cure. This disorder is called long COVID. This paper presents a study carried out at the University Hospital of Ourense with the aim of establishing a relationship among the disease and variations in Heart rate variability (HRV) parameters using machine learning (ML). Five heart rate recordings were obtained per subject, both at rest and under conditions of physical effort and stress. Each record was processed and 15 HRV indices were extracted, giving 75 features per patient. Of these features, 16 were selected to train 10 different ML models: Support Vector Classification, Linear Support Vector Classification, Logistic Regression, Linear Discriminant Analysis, Stochastic Gradient Descent, Multiple Layer Perceptron, Naive Bayes, Random Forest, and Gradient and ADA Boost Classifiers. Results show that the best model, Gradient Boost, achieves an accuracy of 85.2\%, F1-score of 84.9\%, and an Area Under the Receiver Operating Characteristic Curve (AUC) of 0.907, and that all models exceed 0.833 AUC. This study demonstrates an association between long COVID and heart rate variability (HRV), highlighting the utility of machine learning models in identifying this relationship and supporting its diagnose.
\end{abstract}


\section{Introduction}

In the last years, the world has been brutally shocked by the Coronavirus disease (COVID-19) pandemic. According to the World Health Organization, there has been more than 700 million of confirmed cases of COVID-19-infected people, and nearly 7 millions of deaths (\url{https://covid19.who.int/}). 

Patients with COVID-19 presented varied clinical manifestations: dyspnea, muscle pain, sore throat or headache, among others \cite{weng2021pain}. The SARS-CoV-2 virus infects the respiratory system \cite{weng2021pain} but also alters the autonomous nervous system (ANS) \cite{pan2021alteration}.

The ANS controls the heart rate (HR) through its sympathetic (S) and parasympathetic (PS) branches. Increases in the HR are related to sympathetic activity, while reduction in the HR is associated with increased parasympathetic activity. A non-invasive measure of the ANS function is the Heart Rate Variability (HRV), which refers to alterations, beat to beat, in the heart rhythm. The HRV is normally studied from the values of the distances between consecutive beats or RR intervals \cite{malik1990heart}. The clinical usefulness of HRV to detect or deal with different pathologies has been highlighted in many medical areas such as myocardial infarction \cite{malik1989heart}, stress \cite{kim2018stress}, transplantation \cite{dolgoff2012effect} or diabetes \cite{malpas1990heart}. 

Analysis of HRV may provide a quantitative analysis and evaluation of the neurovegetative nervous system. For this task, both linear (time and frequency domain) and nonlinear HRV parameters (see table \ref{tab:HRVGeneralIndices}) can be used \cite{martinez2017}, \cite{shaffer2017overview}.

\begin{table}[h!]
\centering
\renewcommand{\arraystretch}{0.8}
\centering
{\fontsize{10}{20}\selectfont
\begin{tabular}{l}
\hline
\textbf{Frequency domain}\\
    \hspace{0.5cm} $HF$: components in the 0.15 to 0.4 Hz range, related to PS branch (Hz)\\
    \hspace{0.5cm} $HF/LF$ (global): ratio between HF and LF (dimensionless)\\
    \hspace{0.5cm} $HRV$ (global): total spectral power, sum of the power across all the\\
        \hspace{1.2cm} spectrum (Hz$^{2}$)\\
    \hspace{0.5cm} $mHR$: mean value of the HR (bpm)\\
\hline
\textbf{Time domain}\\
    \hspace{0.5cm} $SDNN$ (global): standard deviation of normal RR intervals (ms) \\
    \hspace{0.5cm} $SDSD$ (HF): standard deviation of differences between adjacent RR\\
        \hspace{1.2cm} intervals (ms) \\
    \hspace{0.5cm} $IRRR$ (global): difference between 3rd and 1st quartile (ms)\\
    \hspace{0.5cm} $HRVi$ (LF): HRV triangular index: integral of the intervals histogram\\
        \hspace{1.2cm} divided by its maximum (ms) \\
    \hspace{0.5cm} $pNN50$ (HF): proportion of adjacent RR intervals differing by more\\
        \hspace{1.2cm} than 50 ms (\%)\\
    \hspace{0.5cm} $rMSSD$ (HF): square root of the mean of the squares of differences\\
        \hspace{1.2cm} between adjacent RR intervals (ms) \\
    \hspace{0.5cm} $TINN$ (LF): baseline width of the RR interval histogram (ms) \\
    \hspace{0.5cm} $MADRR$ (HF): median of the absolute differences between adjacent\\
        \hspace{1.2cm} RR intervals\\
\hline
\textbf{Non-linear estimators}\\
    \hspace{0.5cm} $SD1$ (HF): standard deviation of Poincaré plot perpendicular to the\\
        \hspace{1.2cm} the identity line, parasympathetic activity (ms) \\
    \hspace{0.5cm} $SD2$ (LF): standard deviation of the Poincaré plot along the identity\\
        \hspace{1.2cm} line, global variability (ms) \\
    
\hline
\end{tabular}
\caption{Linear and non-linear parameters for HRV analysis. Each parameter includes its category (LF, HF or global), the definition and the units}\label{tab:HRVGeneralIndices}
}
\end{table}

It has been demonstrated that HRV is related to COVID-19. In fact, the SARS-CoV-2 virus can be detected by HRV analysis a few days prior to the presence of the clinical signs of the disease \cite{hirten2021use}, \cite{mason2022detection}. Changes in respiratory rate can also predict the risk of COVID-19 infection \cite{miller2020analyzing}.

The study of HRV and the SARS-CoV-2 virus is not only limited to early detection and prediction. For example, Mol et al. tried to predict mortality and intensive care unit (ICU) referral employing HRV markers in patients hospitalized with COVID-19, and they found that HRV values below median predicted ICU referral within the first week of hospitalization \cite{mol2021heart}. Ponomarev et al. examined the cardiac autonomic responses, by measuring HRV parameters before, after, and during coronavirus disease; even though no statistically significant differences were found in the group analysis, statistically significant changes were found for some individuals during their coronavirus infection \cite{ponomarev2021heart}. Aliani et al. employed HRV parameters to assess the  severity of the disease, reaching a 94\% of accuracy when comparing healthy and minor severity COVID-19 group \cite{aliani2023automatic}. Kaliyaperumal et al. assessed HRV in patients, comparing between subjects suffering from COVID-19 and healthy controls, finding an increased parasympathetic tone in SARS-CoV-2 virus-infected patients \cite{kaliyaperumal2021characterization}. Martínez et al. developed a model for early detection of COVID-19, obtaining a sensitivity score of 0.752, a specificity score of 0.609, being the area under the Receiver Operating Characteristic (ROC) curve AUC of 0.728 \cite{martinez2021machine}.

Individuals suffering from long COVID deserve special attention. These patients can be considered as those who manifest persistent symptoms of the disease, or present chronic or acute sequelae 3 months after the initial SARS-CoV-2 infection. Estimations indicate that around 10–20\% of the virus infected individuals may develop long COVID \cite{who2022longcovid}.

Some effects of long COVID over the ANS have been observed, such  as a sympathetic excitation influence and parasympathetic reduction \cite{marques2022reduction}, or dysautonomia characterized by dysregulation of the HRV \cite{barizien2021clinical}, \cite{mooren2023}. Major efforts are being made to deal with different aspects of long COVID related to HRV. For example, Shah et al., employed HRV time-domain features as markers for cardiovascular dysautonomia, comprising postural orthostatic tachycardia syndrome (POTS) and orthostatic hypotension (OH). Results indicated that the most differential feature was the RMSSD, found to be significantly lower on post COVID patients. Best results showed an AUC of 89.8\% and 89\% accuracy in classification \cite{shah2022}. Menezes et al. found lower values for HF parameters in patients suffering from persistent SARS-CoV-2 infection, compared to healthy individuals \cite{menezes2022}.

The work by Acanfora et al.  related long COVID symptoms with the disfunctionality of the autonomic nervous system through the study of several variables, such as D-Dimer, NT-ProBNP and analysing HRV both on time and frequency domain \cite{acanfora2022}. This study was performed on 50 patients, 30 of them suffering from long COVID. The results showed a significant correlation between long COVID and high frequency components, as well as the LF/HF ratio. The power spectral analysis appears to be lower in long COVID cases, and a significant increase on LF/HF quotient was observed. Recently, Gonçalves et al. found that long COVID reduced HRV at rest and during deep breathing \cite{da2023}, and results obtained by Neves et al. showed a reduction in parasympathetic activation during long COVID, as well as an increase in body temperature due to possible endothelial damage \cite{neves2023}. 

In the last years, Machine Learning (ML) has been rapidly evolving due to its capacity to mimic humans' intelligence by learning from their environment and from experience \cite{el2015machine}. Many different ML models can be found in the literature, such as Support Vector Machine (SVM) \cite{noble2006}, Logistic Regression (LR) \cite{hilbe2009}, Naive Bayes (NB) \cite{berrar2018}, Linear Discriminant Analysis (LDA) \cite{balakrishnama1998}, Random Forest (RF) \cite{biau2012}, Stochastic Gradient Descent  (SGD) \cite{raff2022}, Multiple Layer Perceptron (MLP) \cite{hackeling2017} and Gradient (GB) and ADA (ADA) Boost Classifiers \cite{bahad2020}. Application of ML to different research fields is continuously increasing, and promising results are being obtained in many medical research fields, in combination with HRV: cancer detection \cite{vigier2021cancer}, diabetes prediction \cite{alkhodari2021screening}, or cardiovascular disease \cite{udawat2022automated}. Applied to the SARS-CoV-2 virus, only a few studies employ artificial intelligence to deal with both COVID-19 and long COVID patients via HRV \cite{martinez2021machine}, \cite{shah2022}.

The main goal of the present work was to go one step further in the long COVID research, and to assess its relationship with HRV employing ML models. Specifically, HRV estimators were analyzed in both long COVID patients and controls, trying to establish a relationship between the disease and variations in these HRV parameters. HR data were recorded not only in static conditions, but also in dynamic situations and under certain stress settings.

The rest of the paper is organized as follows: the dataset, feature extraction and processing are described in next section, as well as training and validation methods. Then, section \ref{sec:resdiscuss} presents the major results and discussion. Finally, section \ref{sec:concl_fut} establishes the main conclusions and presents some guidelines to continue the work presented in this paper.

\section{Materials and methods}
\label{sec:matmeth}

\subsection{Study population}

For this work, subjects suffering from persistent COVID-19 and control individuals (fully cured COVID patients) were included. Recruitment was carried out at the University Hospital of Ourense and at the health centers in its area of influence. Ethical approval by the  Clinical Research Ethics Committe from Galicia was obtained. The study was also conducted in accordance with the ethical principles of the Declaration of Helsinki, and registered in ClinicalTrials.gov (ID NCT05629793). Informed consent was provided for all individuals, who had capacity to accept to participate in the study. Inclusion and exclusion criteria for both groups are now detailed.

\begin{itemize}
\item Long COVID group:
 \begin{itemize}
     \item Inclusion criteria: individuals aged between 18 and 70 years old, with confirmed coronavirus infection, and symptoms such as fatigue, respiratory distress or cognitive dysfunction, for more than 2 months.
     \item Exclusion criteria: active COVID-19 infection, with cardiac arrhythmia or pacemaker, Raynaud's disease, or with other types of diseases that impair exercise capacity: uncontrolled congestive heart failure, severe aortic stenosis, pulmonary edema, recent pulmonary thromboembolism, thyrotoxicosis, among others.
 \end{itemize}

 Initially, 66 individuals (44 women, 22 men) (47.71 $\pm$ 9.48 years old) and a mean Body Mass Index (BMI) of 27.90 $\pm$ 5.42 were recruited.

\item Control group:
\begin{itemize}
     \item Inclusion criteria: individuals aged between 18 and 70 years old, with confirmed SARS-CoV-2 infection for some time, but complete functional recovery. In addition, non sequelae nor symptoms existed at least 3 months of having overcome the disease.
     \item Exclusion criteria: the same criteria than in long COVID patients.
 \end{itemize}
 
At first, 69 subjects (40 women, 29 men), with an age of 47.50 $\pm$ 8.08 years and a BMI of 25.18 $\pm$ 4.29 were enrolled in the study.
\end{itemize}

\subsection{Protocol for data collection}

Data were collected in identical conditions for all participants.
All subjects underwent a battery of tests and physiological measurements, and a detailed explanation about the process was given to each one. In order to obtain HRV indexes both at rest and in varied conditions of physical or emotional stress, HR data were recorded in different situations:
\begin{itemize}
    \item \textit{Baseline record (BLR)}: a 3-minute long record at sitting position, and without performing any physical or mental effort.
    \item \textit{Six-minute walking test (6WT)}: this standard in cardiac rehabilitation can be used as a measure of the individual's functional significance \cite{bellet2012}.  Patients were asked to cover the maximum distance by walking straight for 6 minutes \cite{enright2003}, while recording the corresponding HR signal. After performing the walk, a new recovery 2-minute test was acquired.
    \item \textit{Cold pressure test (CPT)}: employed in social Psychology to induce stress provoking pain \cite{bali2015clinical} which produces, as a response, an increase in skin conductance. We used it to reveal differences in sudomotor response between groups. Sudomotor dysfunction has been previously described in Long- Covid in resting conditions \cite{hinduja2021sudomotor}.  This is a simple test consisting in the immersion of the subject's hand in cold water during 1 minute \cite{silverthorn2013}, while she/he stays in prone position. Two records were obtained, one while the hand was in water, and the other during the next 2 minutes.
\end{itemize}

\subsection{HRV data collection}

To obtain the HR records, an modified version of the VARSE open source application was used \cite{cuesta2022}. Each participant was wearing a Bluetooth Polar H10 chest strap\footnote{\url{https://www.polar.com/en/sensors/h10-heart-rate-sensor}} connected to a laptop, placed over his/her chest. The acquisition time varied for each recording, in accordance with the specifications previously indicated. 

Values of the number of beats, mean HR and mean RR were checked for consistency. The process was repeated if an error occurred during acquisition. In total, records for 66 long COVID patients and 69 control subjects were obtained.

\subsection{HR signal processing and feature extraction}

Previously to the data processing, some records were clipped, discarding in all of them the initial segment, prior to the stabilization of the subject \cite{vila2019}. For each subject, we will have:
\begin{itemize}
    \item Baseline record \textbf{BLR}: the 2-minute central fragment, discarding the initial 20 seconds.
    \item From the six-minute walking test (6WT):
    \begin{itemize}
        \item \textbf{6WT}: from the 6-minute record, the 3-minute central portion.
        \item \textbf{6WA}: from the recovery 2-minute record, the central 1-minute segment.
    \end{itemize}
    \item From the cold pressure test (CPT):
    \begin{itemize}
        \item \textbf{CPT}: a 50-second fragment (discarding the first 10 seconds).
        \item \textbf{CPA}: the central 100 seconds portion from the 2-minute record (after removing the hand from cold water).
    \end{itemize}
\end{itemize}

\begin{figure}[htbp]
   \centering
   \includegraphics[trim={0 0 0 0},width=\textwidth]{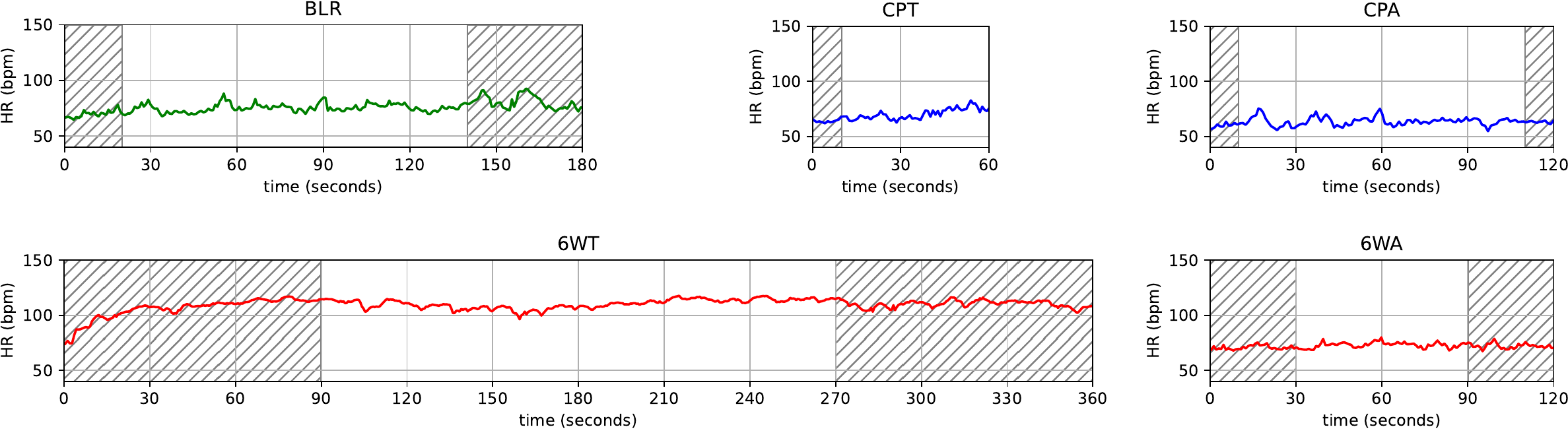}
   \caption{Example of data obtained from a control subject}
   \label{fig:dataRR}
\end{figure}

Then, for each subject we will have five files: BLR, 6WT, 6WA, CPT and CPA. Figure \ref{fig:dataRR} shows an example of these files acquired for a control case.

The RHRV package \cite{martinez2017} was the used to process these records and to obtain the HRV parameters. This process involved the following steps:
\begin{enumerate}
    \item \textit{Signal filtering}: each record was automatically filtered to reduce incorrect data and remove artifacts. \cite{vila1997}.
    \item \textit{Manual filtering}: to remove remaining artifacts from the previous step, in order to obtain good quality records. Recordings from 3 long COVID patients and 3 control group individuals were discarded due to quality issues. 
    \item \textit{Interpolation}: all RR signals were linearly interpolated at 4 Hz.
    \item {\textit{HRV analysis}}: a FFT (Fast Fourier Transform) was employed, using 40-second Hamming windows with 1 second shift. Time and frequency, as well as nonlinear estimators, were extracted.
    \item \textit{Estimation of HRV indexes}: the parameters listed in Table \ref{tab:HRVGeneralIndices} were obtained for each of the recordings of the 129 valid cases (63 long COVID and 66 control subjects).
\end{enumerate}

\subsection{Feature selection}

Once all the HRV estimators were extracted for all the recordings, an initial study was performed to choose which indices provide better discrimination ability between long COVID patients and the control group. This study was performed on features extracted from BLR and 6WT, since they are the largest records and the ones that represent daily situations (basal conditions and physical stress, respectively). 

No statistical evidence was found when analyzing BLR between both groups. However, when some physical effort is exerted by the subjects (6WT), differences between long COVID patients and controls arise, except for the LF/HF ratio. Thus, HRV indexes obtained during the 6WT were considered for the study, with the exception of the LF/HF ratio. 

For each subject, differences in HRV parameters between the BLR and the 6WT (6-B), and between the 6WT and the recovery recording (A-6) were considered to normalize the 6WT values and to include a measure of the recovering capacity. Besides, in order to use information about HRV indices in mental stress situations, the HRV parameters from the CPT situation were also included.

The set of parameters was further reduced with two goals in mind: (i) to remove highly-correlated variables, which could lead to an inefficient and noisy model, and (ii) to keep an adequate and sufficient set of parameters in which both HF and LF categories were represented. The following HRV indexes were then included:
\begin{enumerate}
    \item  rMSSD (HF-related): this parameter proved to be relevant in the exploratory analysis, and it is widely used in the literature regarding HRV and COVID \cite{shah2022}.
    \item HRVi (LF-related): it also yielded statistically significant differences in the exploratory analysis, and it a well-known index used in previous research works \cite{vila2019}.
    \item SD2 (LF-related): it showed overall low correlation with all the other indexes.
    \item mHR: not a pure HRV feature, but an important physiological marker that should be taken into account.
\end{enumerate}

\begin{table}[htbp]
    \centering
    \renewcommand{\arraystretch}{1.2}
    \begin{tabular}{cccccc} 
    \multicolumn{2}{c}{LF-related} & & HF-related & & Physiol. \\
    \cline{1-2} \cline{4-4} \cline{6-6} 
    HRVi(6WT) & SD2(6WT) & & rMSSD(6WT) & & mHR(6WT) \\
    HRVi(6-B) & SD2(6-B) & & rMSSD(6-B) & & mHR(6-B) \\
    HRVi(A-6) & SD2(A-6) & & rMSSD(A-6) & & mHR(A-6) \\
    HRVi(CPT) & SD2(CPT) & & rMSSD(CPT) & & mHR(CPT) \\
    \hline
    \end{tabular}
    \caption{Set of parameters used in our study. Each parameter was considered in four situations: 6WT (effort), 6-B (effort - basal), A-6 (recovery after effort) and CPT (mental stress)}
    \label{tab:indices}
\end{table}

\begin{figure}[htbp]
    \centering
        \includegraphics[width=\textwidth]{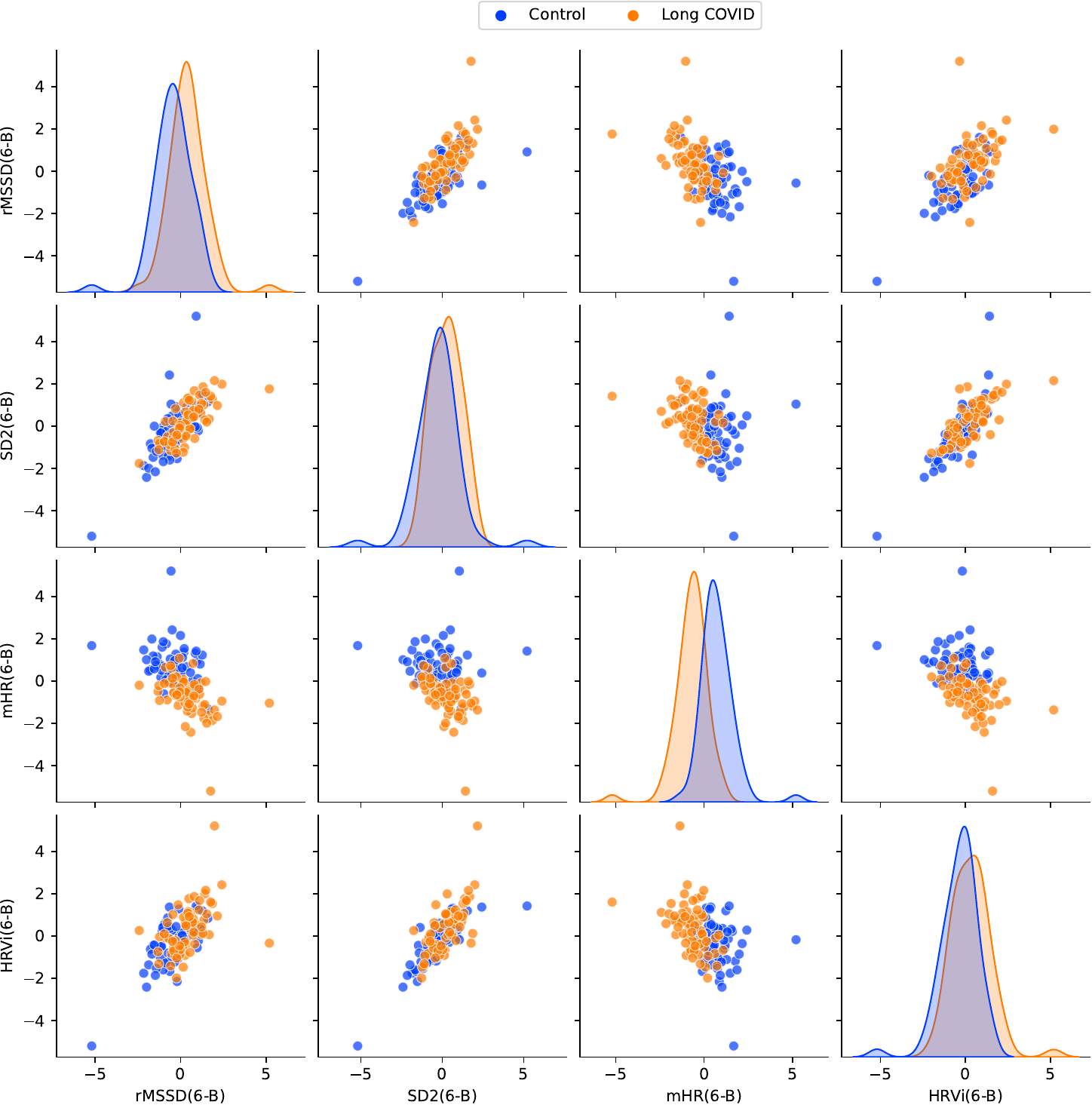}
       \caption{Comparative plots of 6-B records for rMSSD, HRVi, SD2 and mHR parameters}
       \label{fig:baseline_6mwt_diff}
    
\end{figure}

It can be expected that the most prominent differences between Long COVID patients and control subjects appear in the recovery capacity after an effort. However, fig. \ref{fig:baseline_6mwt_diff} compare the distributions of these four parameters for the 6-B records. It can be seen that there are no significant differences except only in mHR, which could be expected as Long COVID patients have a worse physical condition. This fact reinforces the idea of the application of complex machine learning models since the classical statistic analysis does not seem to be enough for the aim of this study.

\begin{figure}[htbp]
   \centering
   \includegraphics[width=\textwidth]{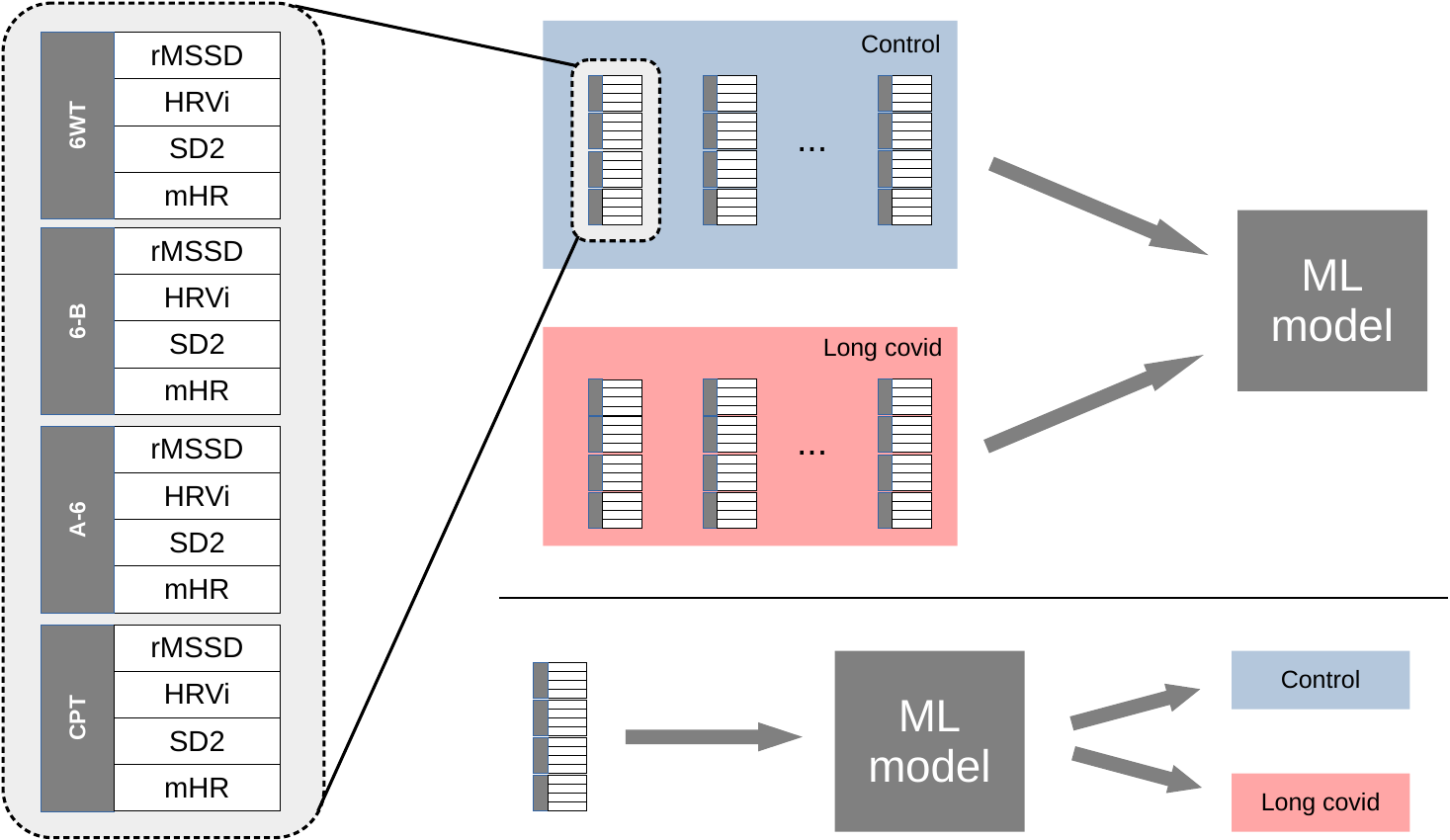}
   \caption{Training and validation procedure}
   \label{fig:trainval}
\end{figure}

The set of parameters used in this study is summarized in table \ref{tab:indices}. As depicted in fig \ref{fig:trainval}, ML models were trained with a collection of 4 features (rMSSD, HRVi, SD2 and mHR), obtained from 4 different sets of records (6WT, 6-B, A-6 and CPT), which resulted in a total of sixteen final features. All parameters were previously normalized, since this transformation has been proven to have a positive influence in ML problems \cite{raju2020}.

\subsection{Training and validation}

The aim of this study is to determine the feasibility of detecting long COVID patients using HRV with ML models (see fig. \ref{fig:trainval}). For this task, the \textit{scikit-learn} Python's module \cite{raschka2020} was used. For each ML approach, a process of hyper-parameterization was first performed where values were selected according to their meaning and following guidelines from the package documentation and publications \cite{agrawal2021}. Besides, all parameters where tested in order to find the most appropriate values. A built-in 5-fold cross validation method offered by \textit{scikit-learn} was also used.  The process is explained in detail on the following paragraphs: 

\begin{itemize}
\item \textbf{Support Vector Machines} (SVMs): this approach operates in high dimensional spaces, aiming to create a clear separation boundary, called a hyperplane, between data points belonging to different categories. For this purpose, 
considering $x_i \in \mathbb{R}^p, i=1,\ldots n$ the set of features and $y \in (1,-1)^n$ the classes, the goal of SVM algorithm is to minimize $\frac{1}{2}\omega^T \omega + C \sum_{1=1}^{n} s_i$ with the restriction $y_i(\omega^T \phi(x_i) +b) \geq 1-s_i, s_i \geq 0, i=1 \ldots n$. The function $\phi()$ is known as the \textit{kernel}.

The \textit{scikit-learn} package offers the possibility of using different kernels. Four different kernels were used: \textit{linear, polynomial, rbf and sigmoid}. In our tests, the set of possible values for the penalty term $C$ went from 0.1 to 0.5 with 0.1 increment. This method is referred as SVC (Support Vector Classifier) in the rest of this paper.

The \textit{scikit-learn} package also offers a specific function that uses a linear kernel, the so-called \textit{LSVC} (Linear Support Vector Classification). In this case it is possible to use the \textit{hinge} and the \textit{square hinge} approaches for the loss function. Both possibilities were tested as well.

\item \textbf{Logistic Regression} (LR): like linear regression, logistic regression builds a model to predict an outcome, but instead of continuous values, it tries to predict the probability of a event by means of a function. For the algorithm used in the optimization problem, also known as the solver, \textit{lbfgs} and \textit{liblinear} were tested. The cost function to be minimized is $C \sum_{1=1}^{n} s_i (-y_i \log (\hat{p}(X_i)) - (1-y_i) \log (1-\hat{p (X_i)})) + r(\omega)$, where the vector $s$ is formed by element-wise multiplication of the class weights and sample weights. The regularization strength $C$ was tuned in the range 0 to 5 with 0.1 shift. The probability of the positive class is given by $\hat{p(X_i)=P(y_i=1|X_i)}$. The regulatory term $r(\omega$) was fixed to the $l_2$ norm ($\frac{1}{2}\omega^T \omega$), aiming at introducing a penalization variable to the input features both to reduce overfitting and improve generalization. 

\item \textbf{Linear Discriminant Analysis} (LDA): the core idea of this approach is to try to find a linear combination of features to best separate different classes in the data. For this purpose, each each sample $X$ is assigned to the classs $k$ that maximize the log-posterior probability $logP(y=k|X)$. The function to be maximized is reduced to $-\frac{1}{2} (x-\mu_k)^t \sum^{-1}(x-\mu_k) +log P(y=k) +C$, where $\sum$ is the covariance matrix, $\mu_k$ is the mean of class $k$ and $C=P(X)=\sum_kP(X|y=k)P(y=k)$. To compute $P(y=k)$ solvers \textit{svd} and \textit{lsqr} were tested.

\item \textbf{Stochatic Gradient Descent} (SGD): the base of this algorithm is to iteratively adjust some parameters to minimize a cost function. In practice, the function to minimize is $E(\omega,b)=\frac{1}{n} \sum_{i=1}^{n}L(y_i,f(x_i)) + \alpha R(\omega)$, being $L$ a loss function that measures the model misfit. The \textit{perceptron} function $L(y_i,f(x_i))=max(0,-y_i f(x_i))$ was used. The parameter that controls the regularization strength ($\alpha>0$), was tried with values 0.0001, 0.001, 0.01, 0.1, 1 and 10. Finally, $R(\omega)=||\omega||^2$, the $L2$ norm, was selected. Finally, the fitting intercept option, which can be True or False, was also tested.

\item \textbf{Multi-Layer Perceptron} (MLP): in this case we used a three-layer neural network. Neural Networks are a very popular ML approach these days, but their main drawback is that they need a big amount of training material to be a successful estimator.

Mathematically, this network can be represented by the function $f(x)=W_2 g(W_1^T x + b_1) + b_2$ where $W_1$ and $W_2$ represent the weights of the input and hidden layer, $g()$ is the activation function and $b_1$ and $b_2$ the bias added of the hidden and output layer. These parameters are tuned to minimize the lost function $loss(\hat{y_i},y_i,W) = \frac{1}{n} \sum_{i=0}^n (y_i \ln{\hat{y_i}}+(1-y_i)\ln({1-\hat{y_i}})) + \frac{\alpha}{2n} || W ||_2^2 $. Four different activation functions (\textit{identity, logistic, tanh} and \textit{relu}) and three solvers (\textit{lbfgs, sgd} and \textit{adam}) were tested. For the number of neurons in the hidden layer, we tried values from 100 to 200 with 25 increment.

\item \textbf{Naive Bayes} (NB):  is a classification algorithm that works by applying Bayes' theorem assuming that the features are statistically independent. The class $y$ is estimated using the following equation: $\hat{y} = \arg\!\max_{y} P(y) \prod_{i=1}^{n} P(x_i | y)$. It also assumes that characteristics follow a Gaussian approximation, that is, $\prod_{i=1}^{n} P(x_i | y) = \frac{1}{\sqrt{2 \pi \sigma_y^2 }} \exp\left( {-\frac{(x_i-\mu_i)^2}{2\sigma_y^2}} \right) $. In this case, no tuning was required.

\item \textbf{Random Forest} (RF): a decision tree is a hierarchical collection of rules that is applied to the features to decide the class. Since an individual tree may present many problems (overfitting, instability), it is better to apply a collection of different trees, using subsets of samples and features, and globally select the class by means of a voting procedure. This method is called Random Forest (RF).

Two different loss functions were tried: \textit{gini} and \textit{entropy}. Depth of the tree was tested with values from 1 to 50 with a step of 1. The minimum split and leaf samples both vary from 1 to 10 and the number of estimators from 10 to 100 with a 10 increment. Lastly, the $bootstrap=FALSE$ option was tested (in this case all the training dataset is used for all trees)  and the number of features to consider for the best split was set to $\sqrt{n\_features}$ .



\item \textbf{Gradient Boost} (GB): this is \textit{ensemble} method that combines weak learners  (different trees) sequentially to build a powerful predictor. The number of minimum samples to split a node were tested with values 20 and 50, as well as minimum samples and maximum depth. The number of estimators was between 11 and 100, with 10 increment.The number of features considered for the best split was set to $\sqrt{n\_features}$.

\item \textbf{ADA Boost} (ADA): this \textit{ensemble} method is similar to the immediately preceding one. It is based on applying a sequence of classifiers, but in each iteration each sample receives a weight. Initially, all samples have the same weight $1/N$. In each stage the misclassified samples  increase their weight. The default values were used for all the parameters, with the exception of the learning rate that was tuned with values 0.2, 0.3, 0.4, 0.5, 1 and 10.

\end{itemize}
Table \ref{tab:FinalTuning} shows the final values used for all these methods after the tuning phase.  When any of these methods required random numbers, a fixed seed was used to eliminate one possible source of non-determinism.

\renewcommand{\arraystretch}{1.25}

\begin{table}[htbp]
    \centering
    \begin{tabular}{ll} 
    \hline
    \textbf{Model} & \textbf{Parameters}  \\ \hline
    \textbf{GB} & max\_depth = 20, max\_features = sqrt, min\_samples\_leaf = 20
    \\ & n\_estimators = 11, min\_samples\_split=50\\
    \textbf{RF} & max\_depth = 2, min\_samples\_split = 1, min\_samples\_leaf = 4,\\ & bootstrap = False, max\_features = sqrt, n\_estimators = 70 \\
    \textbf{MLP} & hiddenLayerSizes = 100, activation = identity, solver = adam \\
    \textbf{ADA} & learning\_rate = 0.2 \\
    \textbf{SVC} & kernel = linear, C = 0.2  \\
    \textbf{LSVC} & loss = squared\_hinge, C = 0.1 \\ 
    \textbf{LR} & C = 0.8, solver = liblinear, penalty = l2, fit\_intercept = True \\
    \textbf{LDA} & solver = svd \\ 
    \textbf{SGD} & alpha = 0.001, penalty = l2 \\ 
    \textbf{NB} & - \\ \hline
    \end{tabular}
    \caption{Final values for all ML models obtained in the tuning phase}
    \label{tab:FinalTuning}
\end{table}

Once all models were configured, a 5-fold cross-validation was performed on the complete dataset \cite{berrar2019}, with final results calculated as the mean of all iterations in the cross-validation process. Performance was evaluated using metrics often employed in ML problems \cite{vujovic2021}: (i) Accuracy, or overall correctness of the model (how many positive or negative predictions are correct), (ii) Precision, or how many positive predictions are actually positive, (iii) Recall or sensitivity, that gives the number of detected positives, (iv) F1-score, which is a balance between precision and recall, and (v) AUC, that measures the overall performance of the model. 

\section{Results and discussion} \label{sec:resdiscuss}

The results of all the ML approaches used in this work are included in Table \ref{tab:FinalResults}. The best results were obtained with the GB classifier, achieving a \textit{90.7\%} AUC and an accuracy of \textit{85.2\%}. On the other hand, SVMs are well-known ML algorithms with steady good results on numeric characterization problems. This is corroborated by performance of both SVC and LSVC algorithm, achieving over \textit{80\% accuracy} and \textit{90\% AUC}. In addition, other models that present consistent good results: RF, ADA and MLP yield over \textit{81\%} of accuracy and \textit{87\%} of AUC. 

The performance differences among the various models are not substantial. All models demonstrate the capability to effectively distinguish between the two groups, as evidenced by AUC values exceeding \textit{83\%} in all cases.

\begin{table}[htbp]
    \centering
    \begin{tabular}{cccccc} 
    \hline
    \textbf{Model} & \textbf{Accuracy} & \textbf{Precision} & \textbf{Recall} & \textbf{F1-score} & \textbf{AUC}  \\ \hline
    \textbf{GB} & 0.852 & 0.867 & 0.838 & 0.849 & 0.907 \\
    \textbf{RF} & 0.827 & 0.828 & 0.839 & 0.831 & 0.881  \\
    \textbf{MLP} & 0.811 & 0.807 & 0.856 & 0.820 & 0.900 \\
    \textbf{ADA} & 0.811 & 0.827 & 0.807 & 0.813 & 0.874  \\
    \textbf{SVC} & 0.803 & 0.778 & 0.873 & 0.818 & 0.904 \\
    \textbf{LSVC} & 0.803 & 0.777 & 0.873 & 0.816 & 0.902 \\
    \textbf{LR} & 0.795 & 0.774 & 0.856 & 0.808 & 0.907  \\
    \textbf{LDA} & 0.770 & 0.754 & 0.824 & 0.783 & 0.875  \\
    \textbf{SGD} & 0.738 & 0.752 & 0.744 & 0.741 & 0.833 \\ 
    \textbf{NB} & 0.755 & 0.794 & 0.730 & 0.751 & 0.838\\\hline
    \end{tabular}
    \caption{Final results of discrimination performance for all ML models of the study}
    \label{tab:FinalResults}
\end{table}


Most previous studies on COVID patients used HR recordings obtained at rest. In our preliminary analysis, we observed that recordings obtained at rest exhibited the least pronounced differences between the control and the long COVID groups.  Consequently, we believe that acquiring recordings while subjects underwent physical effort and stress was particularly valuable. These conditions appear to cause more pronounced difference between the two groups. Possibly, this difference is related to various regulatory dysfunctions in long COVID patients.

For this reason, and in order to gain insight on the problem and justify some results, the importance of the features in models \textit{RF} and \textit{LR} was analyzed. This was performed by studying the features' coefficients obtained from these models.

\begin{figure}
   \centering
   \includegraphics[width=\textwidth]{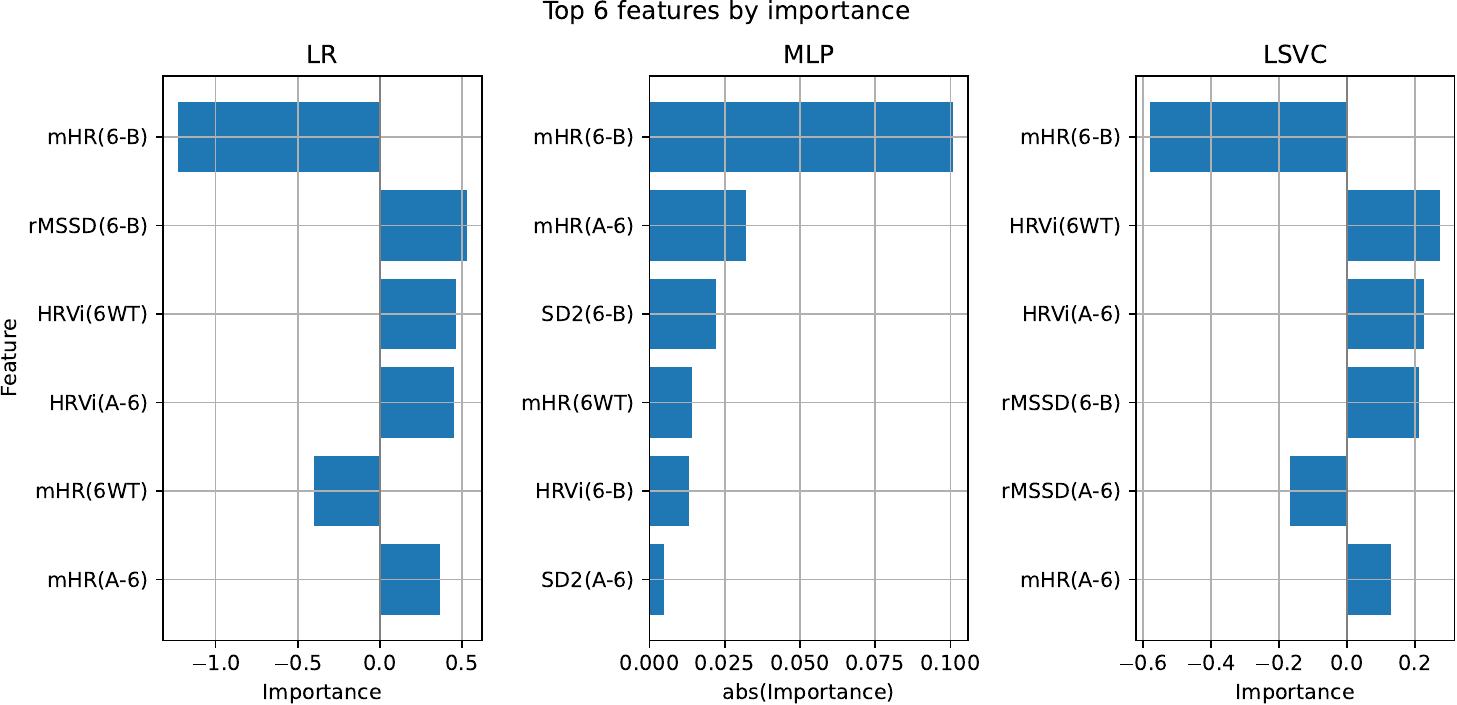}
   \caption{Top relative importances of the features given by the RF model (left) and features with highest log odds ratio magnitude according to the LR model (right)}
   \label{fig:combined}
\end{figure}

The coefficients obtained from the RF model give some interesting results. From the 16 original features, the decision is primarily made by taking 7 of them into account, being the related to mean heart rate the most valuable ones. As it can be seen in the left subplot of fig. \ref{fig:combined}, the three features with the highest impact are all related to the mean heart rate (recovery after effort, during physical effort minus the baseline, and during physical effort). The first pure HRV feature that appears is the rMSSD, which had already being related with post COVID symptoms in  previous works \cite{shah2022}. Features related with mental stress seem to have no significant impact on the model's decision.

In the \textit{LR} model, the parameters used to measure the impact of each feature are the model's \textit{coefficients} which are the log odds ratio for each input, being the most relevant the ones with greater absolute value. The features that seem to have the highest impact on the prediction are \textbf{mHR(6-B)} and \textbf{mHR(6WT)} (see right subplot of fig. \ref{fig:combined}), both related to the mean of the HR. This supports the conclusions obtained from the \textbf{RF} importance features analysis, where the HR related features are also the most discriminating ones.

The analysis of the right subplot of fig. \ref{fig:combined} also reveals some interesting information. Coefficients for \textbf{mHR(6-B)} and \textbf{mHR(6WT)} are negative, which indicates that low values of \textbf{mHR(6-B)} and \textbf{mHR(6WT)} are associated with a higher chance of a record corresponding to a long COVID patient. This may indicate that long COVID affects the sympathetic system, affecting its activation during physically demanding activities. Another interesting fact in this analysis is that features \textbf{SD2} and \textbf{HRVi} (both on CPT conditions) show higher impact on this model than rMSSD, which was the most relevant pure HRV feature in the RF model. 


\section{Conclusions and future work} \label{sec:concl_fut}

In this study, we have tried to assess the capability of HRV measurements to discriminate between long COVID patients and people that has suffered the disease and has made a full recovery. HRV estimators were analyzed and a connection between long COVID and HRV parameters was found.

Other research works that can be found on the literature have studied the impact of COVID on physiological measurements during the illness. However, the number of experiments aiming at establishing comparisons between long COVID and cured people is scarce. Thus, it is difficult to compare the results obtained in this current study against others. However, conclusions of this study can be an interesting starting point to measure and understand the impact on patients suffering from long COVID.

We are not aware of previous works that use HRV and ML techniques to compare long COVID patients against cured COVID subjects. Thus, the results presented in this paper establish an initial step for future work.  

An interesting possibility to continue this study would be to explore different processing techniques, and model selection and tuning. Finally, we think that a very promising way to improve the results would be to combine HRV features with other different physiological markers. 

\section*{Declarations}

\noindent {\bf Funding: } the authors would like to thank the Xunta de Galicia for their financial support in carrying out this work under the project IN852D 2021/20 (co-financed by the European Union with FEDER founds).\\

\noindent {\bf Conflict of interest: } the authors declare that they have no conflict of interest.\\

\noindent {\bf Availability of data and material: } data for the experiments is not publicly available.\\

    \noindent {\bf Authors’ contribution: } Conceptualization, María J. Lado and Xosé A. Vila; methodology, Brais Iglesias-Otero, María J. Lado and Xosé A. Vila; software, Brais Iglesias-Otero, Arturo J. Méndez and Baltasar García Perez‐Schofield; validation, Leandro Rodríguez-Liñares, Pedro Cuesta-Morales and Baltasar García Perez‐Schofield; investigation, María J. Lado, Arturo J. Méndez, Baltasar García Perez‐Schofield, Pedro Cuesta-Morales, Maria Bustillo and Alexandre García-Caballero; writing original draft, Brais Iglesias-Otero, Xosé A. Vila and Alexandre García-Caballero; writing review and editing, María J. Lado, Leandro Rodríguez-Liñares and Arturo J. Méndez; visualization, Leandro Rodríguez-Liñares, Pedro Cuesta-Morales and Baltasar García Perez‐Schofield; supervision, Xosé A. Vila and Alexandre García-Caballero; funding acquisition, María J. Lado, Arturo J. Méndez, Leandro Rodríguez-Liñares, Baltasar García Perez‐Schofield, Pedro Cuesta-Morales and Xosé A. Vila.\\

\noindent{\bf Ethics approval: } all procedures performed in this study involving human participants were in accordance with the ethical standards of the institutional and/or national research committee and with the 1964 Helsinki declaration and its later amendments or comparable ethical standards.\\

\bibliographystyle{unsrtnat}
\bibliography{references}  

@article{weng2021pain,
  title={Pain symptoms in patients with coronavirus disease (COVID-19): A literature review},
  author={Weng, Lin-Man and Su, Xuan and Wang, Xue-Qiang},
  journal={Journal of Pain Research},
  pages={147--159},
  year={2021},
  publisher={Taylor \& Francis}
}

@article{pan2021alteration,
  title={Alteration of autonomic nervous system is associated with severity and outcomes in patients with COVID-19},
  author={Pan, Yuchen and Yu, Zhiyao and Yuan, Yuan and Han, Jiapeng and Wang, Zhuo and Chen, Hui and Wang, Songyun and Wang, Zhen and Hu, Huihui and Zhou, Liping and others},
  journal={Frontiers in physiology},
  volume={12},
  pages={630038},
  year={2021},
  publisher={Frontiers Media SA}
}

@article{bellet2012,
  title={The 6-minute walk test in outpatient cardiac rehabilitation: validity, reliability and responsiveness—a systematic review},
  author={Bellet, R Nicole and Adams, Lewis and Morris, Norman R},
  journal={Physiotherapy},
  volume={98},
  number={4},
  pages={277--286},
  year={2012},
  publisher={Elsevier}
}

@article{mooren2023,
  title={Autonomic dysregulation in long-term patients suffering from Post-COVID-19 Syndrome assessed by heart rate variability},
  author={Mooren, Frank C and B{\"o}ckelmann, Irina and Waranski, Melina and Kotewitsch, Mona and Teschler, Marc and Sch{\"a}fer, Hendrik and Schmitz, Boris},
  journal={Scientific Reports},
  volume={13},
  number={1},
  pages={15814},
  year={2023},
  publisher={Nature Publishing Group UK London}
}

@article{da2023,
  title={Impact of long COVID on the heart rate variability at rest and during deep breathing maneuver},
  author={da Silva, Andr{\'e}a L{\'u}cia Gon{\c{c}}alves and Vieira, Luana dos Passos and Dias, Luiza Scheffer and Prestes, Cec{\'\i}lia Vieira and Back, Guilherme Dionir and Goulart, Cassia da Luz and Arena, Ross and Borghi-Silva, Audrey and Trimer, Renata},
  journal={Scientific Reports},
  volume={13},
  number={1},
  pages={22695},
  year={2023},
  publisher={Nature Publishing Group UK London}
}

@article{menezes2022,
  title={Cardiac Autonomic Function in Long COVID-19 Using Heart Rate Variability: An Observational Cross-Sectional Study},
  author={Menezes Junior, Antonio da Silva and Schr{\"o}der, Aline Andressa and Botelho, Silvia Mar{\c{c}}al and Resende, Aline Lazara},
  journal={Journal of Clinical Medicine},
  volume={12},
  number={1},
  pages={100},
  year={2022},
  publisher={MDPI}
}

@article{neves2023,
  title={Imbalance of Peripheral Temperature, Sympathovagal, and Cytokine Profile in Long COVID},
  author={Neves, Pablo Fabiano Moura das and Quaresma, Juarez Ant{\^o}nio Sim{\~o}es and Queiroz, Maria Alice Freitas and Silva, Camilla Costa and Maia, Enzo Varela and Oliveira, Jo{\~a}o Sergio de Sousa and Neves, Carla Manuela Almeida das and Mendon{\c{c}}a, Suellen da Silva and Falc{\~a}o, Aline Semblano Carreira and Melo, Giovana Salom{\~a}o and others},
  journal={Biology},
  volume={12},
  number={5},
  pages={749},
  year={2023},
  publisher={MDPI}
}

@book{el2015machine,
  title={What is machine learning?},
  author={El Naqa, Issam and Murphy, Martin J},
  year={2015},
  publisher={Springer}
}

@article{vigier2021cancer,
  title={Cancer classification using machine learning and HRV analysis: preliminary evidence from a pilot study},
  author={Vigier, Marta and Vigier, Benjamin and Andritsch, Elisabeth and Schwerdtfeger, Andreas R},
  journal={Scientific Reports},
  volume={11},
  number={1},
  pages={22292},
  year={2021},
  publisher={Nature Publishing Group UK London}
}

@article{alkhodari2021screening,
  title={Screening cardiovascular autonomic neuropathy in diabetic patients with microvascular complications using machine learning: a 24-hour heart rate variability study},
  author={Alkhodari, Mohanad and Rashid, Mamunur and Mukit, Mohammad Abdul and Ahmed, Khawza I and Mostafa, Raqibul and Parveen, Sharmin and Khandoker, Ahsan H},
  journal={IEEE Access},
  volume={9},
  pages={119171--119187},
  year={2021},
  publisher={IEEE}
}

@article{udawat2022automated,
  title={An automated detection of atrial fibrillation from single-lead ECG using HRV features and machine learning},
  author={Udawat, Abhimanyu Singh and Singh, Pushpendra},
  journal={Journal of Electrocardiology},
  volume={75},
  pages={70--81},
  year={2022},
  publisher={Elsevier}
}

@article{shaffer2017overview,
  title={An overview of heart rate variability metrics and norms},
  author={Shaffer, Fred and Ginsberg, Jay P},
  journal={Frontiers in public health},
  pages={258},
  year={2017},
  publisher={Frontiers}
}

@article{malpas1990heart,
  title={Heart-rate variability and cardiac autonomic function in diabetes},
  author={Malpas, Simon C and Maling, Timothy JB},
  journal={Diabetes},
  volume={39},
  number={10},
  pages={1177--1181},
  year={1990},
  publisher={Am Diabetes Assoc}
}

@article{malik1990heart,
  title={Heart rate variability},
  author={Malik, Marek and Camm, A John},
  journal={Clinical cardiology},
  volume={13},
  number={8},
  pages={570--576},
  year={1990},
  publisher={Wiley Online Library}
}

@article{malik1989heart,
  title={Heart rate variability in relation to prognosis after myocardial infarction: selection of optimal processing techniques},
  author={Malik, Marek and Farrell, T and Cripps, T and Camm, AJ},
  journal={European heart journal},
  volume={10},
  number={12},
  pages={1060--1074},
  year={1989},
  publisher={Oxford University Press}
}

@article{kim2018stress,
  title={Stress and heart rate variability: a meta-analysis and review of the literature},
  author={Kim, Hye-Geum and Cheon, Eun-Jin and Bai, Dai-Seg and Lee, Young Hwan and Koo, Bon-Hoon},
  journal={Psychiatry investigation},
  volume={15},
  number={3},
  pages={235},
  year={2018},
  publisher={Korean Neuropsychiatric Association}
}

@article{dolgoff2012effect,
  title={Effect of laughter yoga on mood and heart rate variability in patients awaiting organ transplantation: a pilot study},
  author={Dolgoff-Kaspar, Rima and Baldwin, Ann and Johnson, M Scott and Edling, Nancy and Sethi, Gulshan K},
  journal={Alternative therapies},
  volume={18},
  number={4},
  pages={53--58},
  year={2012}
}

@article{mol2021heart,
  title={Heart-rate-variability (HRV), predicts outcomes in COVID-19},
  author={Mol, Maartje BA and Strous, Maud TA and van Osch, Frits HM and Vogelaar, F Jeroen and Barten, Dennis G and Farchi, Moshe and Foudraine, Norbert A and Gidron, Yori},
  journal={PLoS One},
  volume={16},
  number={10},
  pages={e0258841},
  year={2021},
  publisher={Public Library of Science San Francisco, CA USA}
}

@article{ponomarev2021heart,
  title={Heart rate variability as a prospective predictor of early Covid-19 symptoms},
  author={Ponomarev, Alexey and Tyapochkin, Konstantin and Surkova, Ekaterina and Smorodnikova, Evgeniya and Pravdin, Pavel},
  journal={medRxiv},
  pages={2021--07},
  year={2021},
  publisher={Cold Spring Harbor Laboratory Press}
}

@article{aliani2023automatic,
  title={Automatic COVID-19 severity assessment from HRV},
  author={Aliani, Cosimo and Rossi, Eva and Luchini, Marco and Calamai, Italo and Deodati, Rossella and Spina, Rosario and Francia, Piergiorgio and Lanata, Antonio and Bocchi, Leonardo},
  journal={Scientific Reports},
  volume={13},
  number={1},
  pages={1713},
  year={2023},
  publisher={Nature Publishing Group UK London}
}

@article{kaliyaperumal2021characterization,
  title={Characterization of cardiac autonomic function in COVID-19 using heart rate variability: A hospital based preliminary observational study},
  author={Kaliyaperumal, Deepalakshmi and Rk, Karthikeyan and Alagesan, Murali and Ramalingam, Sudha},
  journal={Journal of Basic and Clinical Physiology and Pharmacology},
  volume={32},
  number={3},
  pages={247--253},
  year={2021},
  publisher={De Gruyter}
}

@article{marques2022reduction,
  title={Reduction of cardiac autonomic modulation and increased sympathetic activity by heart rate variability in patients with long COVID},
  author={Marques, Karina Carvalho and Silva, Camilla Costa and Trindade, Steffany da Silva and Santos, Marcio Clementino de Souza and Rocha, Rodrigo Santiago Barbosa and Vasconcelos, Pedro Fernando da Costa and Quaresma, Juarez Ant{\^o}nio Sim{\~o}es and Falc{\~a}o, Luiz F{\'a}bio Magno},
  journal={Frontiers in cardiovascular medicine},
  volume={9},
  pages={1046},
  year={2022},
  publisher={Frontiers}
}

@article{hirten2021use,
  title={Use of physiological data from a wearable device to identify SARS-CoV-2 infection and symptoms and predict COVID-19 diagnosis: observational study},
  author={Hirten, Robert P and Danieletto, Matteo and Tomalin, Lewis and Choi, Katie Hyewon and Zweig, Micol and Golden, Eddye and Kaur, Sparshdeep and Helmus, Drew and Biello, Anthony and Pyzik, Renata and others},
  journal={Journal of medical Internet research},
  volume={23},
  number={2},
  pages={e26107},
  year={2021},
  publisher={JMIR Publications Inc., Toronto, Canada}
}

@article{mason2022detection,
  title={Detection of COVID-19 using multimodal data from a wearable device: results from the first TemPredict Study},
  author={Mason, Ashley E and Hecht, Frederick M and Davis, Shakti K and Natale, Joseph L and Hartogensis, Wendy and Damaso, Natalie and Claypool, Kajal T and Dilchert, Stephan and Dasgupta, Subhasis and Purawat, Shweta and others},
  journal={Scientific reports},
  volume={12},
  number={1},
  pages={3463},
  year={2022},
  publisher={Nature Publishing Group UK London}
}

@article{barizien2021clinical,
  title={Clinical characterization of dysautonomia in long COVID-19 patients},
  author={Barizien, Nicolas and Le Guen, Morgan and Russel, St{\'e}phanie and Touche, Pauline and Huang, Florent and Vall{\'e}e, Alexandre},
  journal={Scientific reports},
  volume={11},
  number={1},
  pages={1--7},
  year={2021},
  publisher={Springer}
}

@article{miller2020analyzing,
  title={Analyzing changes in respiratory rate to predict the risk of COVID-19 infection},
  author={Miller, Dean J and Capodilupo, John V and Lastella, Michele and Sargent, Charli and Roach, Gregory D and Lee, Victoria H and Capodilupo, Emily R},
  journal={PloS one},
  volume={15},
  number={12},
  pages={e0243693},
  year={2020},
  publisher={Public Library of Science San Francisco, CA USA}
}

@article{martinez2021machine,
  title={A machine learning approach as an aid for early covid-19 detection},
  author={Martinez-Velazquez, Roberto and Tob{\'o}n V, Diana P and Sanchez, Alejandro and El Saddik, Abdulmotaleb and Petriu, Emil},
  journal={Sensors},
  volume={21},
  number={12},
  pages={4202},
  year={2021},
  publisher={MDPI}
}

@article{acanfora2022,
  title={Impaired vagal activity in long-COVID-19 patients},
  author={Acanfora, Domenico and Nolano, Maria and Acanfora, Chiara and Colella, Camillo and Provitera, Vincenzo and Caporaso, Giuseppe and Rodolico, Gabriele Rosario and Bortone, Alessandro Santo and Galasso, Gennaro and Casucci, Gerardo},
  journal={Viruses},
  volume={14},
  number={5},
  pages={1035},
  year={2022},
  publisher={MDPI}
}

@article{shah2022,
  title={Heart rate variability as a marker of cardiovascular dysautonomia in post-COVID-19 syndrome using artificial intelligence},
  author={Shah, Bhushan and Kunal, Shekhar and Bansal, Ankit and Jain, Jayant and Poundrik, Shubhankar and Shetty, Manu Kumar and Batra, Vishal and Chaturvedi, Vivek and Yusuf, Jamal and Mukhopadhyay, Saibal and others},
  journal={Indian pacing and electrophysiology journal},
  volume={22},
  number={2},
  pages={70--76},
  year={2022},
  publisher={Elsevier}
}

@inproceedings{raju2020,
  title={Study the influence of normalization/transformation process on the accuracy of supervised classification},
  author={Raju, VN Ganapathi and Lakshmi, K Prasanna and Jain, Vinod Mahesh and Kalidindi, Archana and Padma, V},
  booktitle={2020 Third International Conference on Smart Systems and Inventive Technology (ICSSIT)},
  pages={729--735},
  year={2020},
  organization={IEEE}
}

@article{raschka2020,
  title={Machine learning in python: Main developments and technology trends in data science, machine learning, and artificial intelligence},
  author={Raschka, Sebastian and Patterson, Joshua and Nolet, Corey},
  journal={Information},
  volume={11},
  number={4},
  pages={193},
  year={2020},
  publisher={Multidisciplinary Digital Publishing Institute}
}

@inbook{berrar2019,
  author={Berrar, Daniel},
  year= 2019, 
  chapter={Cross-Validation}, 
  editor = {S. Ranganathan and M. Gribskov and K. Nakai and C. Schönbach}, 
  title= {Encyclopedia of Bioinformatics and Computational Biology}, 
  publisher= {Elsevier}, 
}

@book{hilbe2009,
  title={Logistic regression models},
  author={Hilbe, Joseph M},
  year={2009},
  publisher={CRC press}
}

@article{noble2006,
  title={What is a support vector machine?},
  author={Noble, William S},
  journal={Nature biotechnology},
  volume={24},
  number={12},
  pages={1565--1567},
  year={2006},
  publisher={Nature Publishing Group UK London}
}

@article{berrar2018,
  title={Bayes’ theorem and naive Bayes classifier},
  author={Berrar, Daniel},
  journal={Encyclopedia of Bioinformatics and Computational Biology: ABC of Bioinformatics},
  volume={403},
  pages={412},
  year={2018},
  publisher={Elsevier Science Publisher Amsterdam, The Netherlands}
}

@article{balakrishnama1998,
  title={Linear discriminant analysis-a brief tutorial},
  author={Balakrishnama, Suresh and Ganapathiraju, Aravind},
  journal={Institute for Signal and information Processing},
  volume={18},
  number={1998},
  pages={1--8},
  year={1998},
  publisher={Mississippi}
}

@article{biau2012,
  title={Analysis of a random forests model},
  author={Biau, G{\'e}rard},
  journal={The Journal of Machine Learning Research},
  volume={13},
  number={1},
  pages={1063--1095},
  year={2012},
  publisher={JMLR. org}
}

@book{raff2022,
  title={Inside deep learning: Math, algorithms, models},
  author={Raff, Edward},
  year={2022},
  publisher={Simon and Schuster}
}

@article{vujovic2021,
  title={Classification model evaluation metrics},
  author={Vujovi{\'c}, {\v{Z}}},
  journal={International Journal of Advanced Computer Science and Applications},
  volume={12},
  number={6},
  pages={599--606},
  year={2021}
}

@article{enright2003,
  title={The six-minute walk test},
  author={Enright, Paul L},
  journal={Respiratory care},
  volume={48},
  number={8},
  pages={783--785},
  year={2003},
  publisher={Respiratory Care}
}

@article{silverthorn2013,
  title={Cold stress and the cold pressor test},
  author={Silverthorn, Dee U and Michael, Joel},
  journal={Advances in Physiology Education},
  volume={37},
  number={1},
  pages={93--96},
  year={2013},
  publisher={American Physiological Society Bethesda, MD}
}

@article{martinez2017,
  title={Heart rate variability analysis with the R package RHRV},
  author={Mart{\'\i}nez, Constantino Antonio Garc{\'\i}a and Quintana, Abraham Otero and Vila, Xos{\'e} A and Touri{\~n}o, Mar{\'\i}a Jos{\'e} Lado and Rodr{\'\i}guez-Li{\~n}ares, Leandro and Presedo, Jes{\'u}s Mar{\'\i}a Rodr{\'\i}guez and Pen{\'\i}n, Arturo Jos{\'e} M{\'e}ndez},
  year={2017},
  publisher={Springer}
}

@article{vila2019,
  title={Evidence based recommendations for designing heart rate variability studies},
  author={Vila, Xos{\'e} A and Lado, Mar{\'\i}a J and Cuesta-Morales, P},
  journal={Journal of Medical Systems},
  volume={43},
  number={10},
  pages={311},
  year={2019},
  publisher={Springer}
}

@article{vila1997,
  title={Time-frequency analysis of heart-rate variability},
  author={Vila, J and Palacios, F and Presedo, J and Fernandez-Delgado, M and Felix, P and Barro, S},
  journal={IEEE Engineering in Medicine and Biology Magazine},
  volume={16},
  number={5},
  pages={119--126},
  year={1997},
  publisher={IEEE}
}

@article{cuesta2022,
  title={VARSE: Android app for real-time acquisition and analysis of heart rate signals},
  author={Cuesta-Morales, Pedro and Perez-Schofield, Baltasar Garc{\'\i}a and Rodr{\'\i}guez-Li{\~n}ares, Leandro and Lado, Mar{\'\i}a J and M{\'e}ndez, Arturo J and Vila, Xos{\'e} A},
  journal={International Journal of Medical Informatics},
  volume={160},
  pages={104692},
  year={2022},
  publisher={Elsevier}
}

@inproceedings{bahad2020,
  title={Study of adaboost and gradient boosting algorithms for predictive analytics},
  author={Bahad, Pritika and Saxena, Preeti},
  booktitle={International Conference on Intelligent Computing and Smart Communication 2019: Proceedings of ICSC 2019},
  pages={235--244},
  year={2020},
  organization={Springer}
}

@book{hackeling2017,
  title={Mastering Machine Learning with scikit-learn},
  author={Hackeling, Gavin},
  year={2017},
  publisher={Packt Publishing Ltd}
}

@article{agrawal2021,
  title={Hyperparameter optimization using scikit-learn},
  author={Agrawal, Tanay and Agrawal, Tanay},
  journal={Hyperparameter optimization in machine learning: make your machine learning and deep learning models more efficient},
  pages={31--51},
  year={2021},
  publisher={Springer}
}

@misc{who2022longcovid,
  title = {{Post COVID-19 condition (Long COVID)}},
  author = {World Health Organization},
  howpublished = "\url{https://www.who.int/europe/news-room/fact-sheets/item/post-covid-19-condition}",
  year = {2022}, 
  note = "[Online; accessed 17-June-2024]"
}

@article{bali2015clinical,
  title={Clinical experimental stress studies: methods and assessment},
  author={Bali, Anjana and Jaggi, Amteshwar Singh},
  journal={Reviews in the Neurosciences},
  volume={26},
  number={5},
  pages={555--579},
  year={2015},
  publisher={De Gruyter}
}

@article{hinduja2021sudomotor,
  title={Sudomotor dysfunction in patients recovered from COVID-19},
  author={Hinduja, Anand and Moutairou, Abdul and Calvet, Jean-Henri},
  journal={Neurophysiologie Clinique},
  volume={51},
  number={2},
  pages={193--196},
  year={2021},
  publisher={Elsevier}
}






\end{document}